\documentclass[prb,preprint,amsmath,amssymb]{revtex4}
\usepackage{color}
\begin{document}
\author{Kevin Leung$^*$}
\affiliation{Sandia National Laboratories, MS 1415, Albuquerque, NM 87185\\
$^*${\tt kleung@sandia.gov}}
\date{\today}
\title{First Principles Determination of the Potential-of-zero-charge in
an Alumina-coated Aluminum/Water Interface Model for Corrosion Applications}

\input epsf.sty
 
\begin{abstract}

The surfaces of most metals immersed in aqueous electrolytes have a
several-nanometer-thick oxide/hydroxide surface layer.
This gives rise to the existence of both metal$|$oxide and oxide$|$liquid
electrotlyte interfaces, and makes it
challenging to correlate atomic length-scale structures with electrochemical
properties such the potential-of-zero-charge (PZC).  The PZC has been
shown to be correlated the pitting onset potential for corrosion.  In this
work, we conduct large-scale Density Functional Theory and {\it ab initio}
molecular dynamics to calculate the PZC of a
Al(111)$|$$\gamma$-Al$_2$O$_3$(110)$|$water double-interface model within the
context of aluminum corrosion.  By partitioning the multiple interfaces
involved into binary components with additive contributions to the overall
work function and voltage, we predict the PZC to be -1.53~V~vs.~SHE for this
model.  We also calculate the orbital energy levels of defects like
oxygen vacancies in the oxide, which are critical parameters
in theories associated with pitting corrosion.  We predict that the
Fermi level at the PZC lies above the impurity defect levels of the oxygen
vacancies, which are therefore uncharged at the PZC.  From the PZC estimate,
we predict the voltage needed to create oxygen vacancies with net positive
charges within a flat-band approximation.

\end{abstract}
 
\maketitle
 
\section{Introduction}
\label{intro}

\begin{figure}
\centerline{\hbox{  \epsfxsize=6.50in \epsfbox{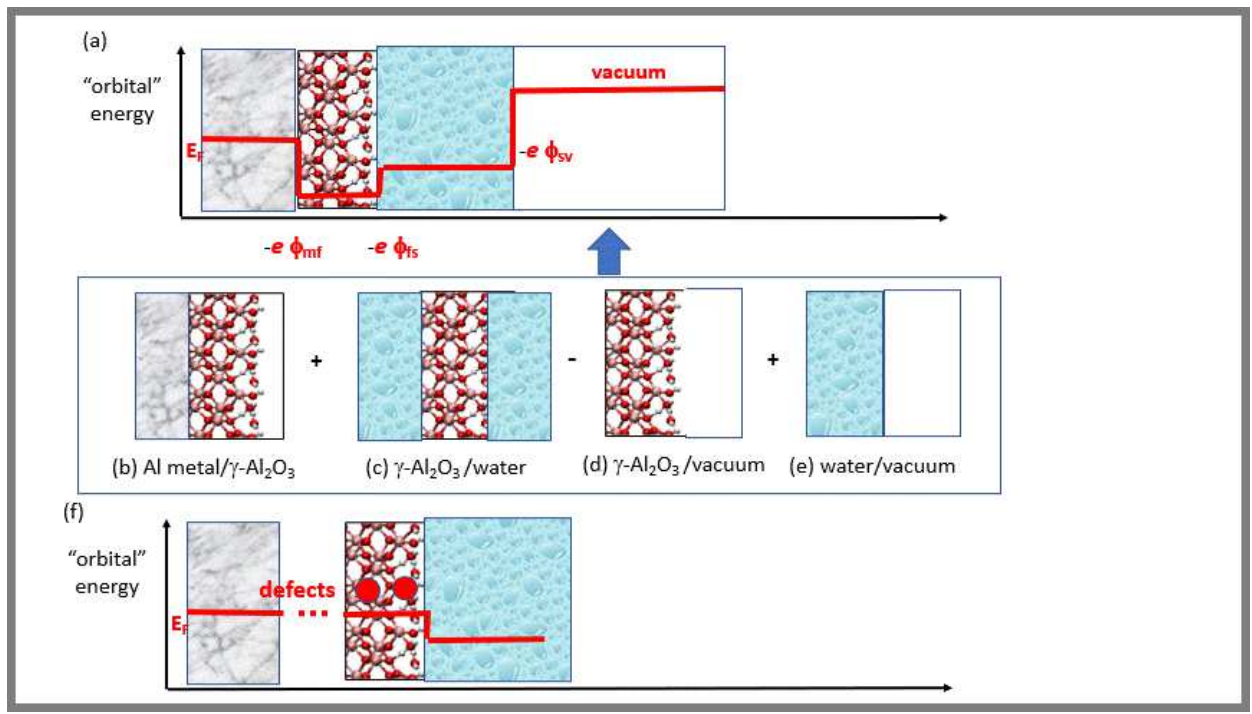} }}
\caption[]
{\label{fig1} \noindent
(a) Schematic of absolute work function for a charge neutral 
Al$|$$\gamma$-Al$_2$O$_3$$|$water.  The work function is related to
the voltage; it can be split into four additive contributions:
(b) Al$|$$\gamma$-Al$_2$O$_3$$|$vacuum;
(c) $\gamma$-Al$_2$O$_3$$|$water; (d) the negative of
$\gamma$-Al$_2$O$_3$$|$vacuum; and (e) water$|$vacuum.
(f) In the experimental literature,\cite{bockris} the metal
is often omitted in the analysis.  Instead, the Fermi level is
assumed to be pinned at defect levels in the oxide, like they
would be in semiconductors.
}
\end{figure}

Corrosion\cite{corrbook} involves complex interfaces and processes that have
so far defied quantitative elucidation using atomic length-scale calculations.
One challenging aspect is the surface-film-covered (non-pristine) nature of 
metal/liquid electrolyte interfaces.  Net charges can exist at the
oxide/electrolyte interface via acid-base reactions at surface hydroxyl
(-OH) sites, compensated by the electric double layer (EDL) in the 
liquid electrolyte immediately outside the passivating oxide
film.\cite{mccafferty10,mccafferty99,natishan88}
Charges can also exist in defect sites in the oxide film itself, or
at the metal$|$oxide interface (Fig.~\ref{fig1}a).\cite{macdonald81a,corros1} 
This complexity makes electrochemical properties like the
potential-of-zero-charge (PZC) far more difficult to predict using
atomic length-scale methods than those for pristine electrodes without
surface films.\cite{juncheng,taylor,otani,pristine} 

The PZC is of significant interest because PZC's of a set of metal alloys
have been demonstrated to be linearly correlated to the pitting (onset)
potentials.\cite{bockris,mccafferty99} It is also correlated to the
pH-of-zero-charge (``pH$_{\rm PZC}$'' , or sometime the ``isoelectric
point'').\cite{mccafferty99,mccafferty10,natishan88} In turn, 
pH$_{\rm PZC}$'s have been correlated with the pH's-of-zero-charge in pure
oxides.\cite{mccafferty10}  From an atomic length-scale computational
viewpoint, the PZC is an equilibrium property which should be amenable to
modeling studies.  At the PZC, metal$|$oxide$|$electrolyte systems exhibit no
overall electric field, and are at the ``flat band'' condition\cite{campbell}
which should the most straight-forward to model.  In contrast, the pitting
potential,\cite{bockris,pittingvoltage} while critical to understanding
corrosion, is a kinetic property, involves extensive chemical reactions,
and its prediction is far beyond the current capability of atomic modeling
methods.

In particular, Density Functional Theory (DFT) predictions of PZC's will help
establish voltage-function relationships which
have been largely lacking in current DFT modeling of explicit metal$|$oxide
interfaces.\cite{marks,costa,costa2,costa3,liu2021,kadowaki,corros1}
For example, Ref.~\onlinecite{corros1} predicted that uncharged oxygen
vacancies (V$_{\rm O}^0$) give way to charged V$_{\rm O}^{2+}$ upon applying a
sufficient electric field via adding a negative charge at the
outer oxide surface.  This is significant because V$_{\rm O}^{2+}$, and not
V$_{\rm O}^0$, are foundational assumptions of the point defect model (PDM) for
corrosion,\cite{macdonald81a,macdonald81b,macdonald16a,macdonald15,macdonald16b} which is one of several theories associated with passivation film
breakdown.\cite{marks,theory1,theory2,theory3,theory4,theory5,theory6,theory7}
However, the voltage associated with this electric field was not elucidated in
Ref.~\onlinecite{corros1}, partly because it was sensitive to the surface
water configuration computed using a sub-monolayer of water frozen at T=0~K.
Therefore the predictions could not be directly compared with measurements
performed at particular voltages.\cite{corros1}  

Here we adopt a bulk-like liquid water boundary condition, using {\it ab initio}
molecular dynamics (AIMD) to treat the aqueous environment (Fig.~\ref{fig1}a),
not monolayers of water at T=0~K.\cite{corros1}   By construction, all
surfaces/interfaces are uncharged.  Thus our models correspond to the PZC,
flat band condition; the electrolyte is pure water and automatically
accounts for the pH-of-zero-charge condition.  We will compute the voltage
vs.~the Standard Hydrogen Electrode (SHE) reference using the work function
approach.\cite{otani12,corros1,pccp} Since it is costly to apply AIMD to
the entire model (Fig.~\ref{fig1}a), we split the interfacial contributions
to the voltage/work function into four parts (Fig.~\ref{fig1}b-e), and
add up the components.  Such a decomposition would have been more
challenging if surface charges and electric fields exist in the oxide film.

Experimentally, Al with anodized oxide films have been estimated to exhibit a
PZC at about -1~V vs.~SHE (Fig.~\ref{fig1}f).\cite{bockris}  A flat band 
potential of $\sim$0.7~V has also been reported;\cite{mccafferty99} the
discrepancy may be due to the oxide film structure and/or experimental
conditions.  The true oxide films associated with these experiments are 
too complex for DFT modeling.\cite{bockris} They have varying Al:O atomic
ratios which suggest a finite oxyhydroxide (e.g., AlOOH) content, and exhibit
negative space charges, which have been
explained in DFT modeling of amorphous oxides.\cite{shluger2,shluger1}  
Instead, we focus on the $\gamma$-phase Al$_2$O$_3$ which has a lower density
and should be a better description of amorphous
Al$_2$O$_3$\cite{persson,shluger2,shluger1} than the
$\alpha$-phase.\cite{corros1}   We compare the DFT-computed PZC
with experimental values, and rationalize the difference in terms of the
model oxide structure.  As the precise atomic length-scale structures
of passivating oxide films and their interfaces are not known experimentally,
we argue that, in the future, we should contruct models that reproduce the
experimental PZC.  In other words, calculated voltages should be treated
as constraints for model construction, and not as ``predictions.''

Finally, we calculate the alignment between a V$_{\rm O}^{0}$ in the oxide
film and the Al metal Fermi level ($E_{\rm F}$) pegged to the PZC.
This alignment allows us to estimate the voltage, offset from
the PZC, at which this particular model exhibits a net charge (i.e.,
becomes V$_{\rm O}^{2+}$) within a flat band approximation, and compare it
with the experimental pitting potential.  \color{black}
We hypothesize that the voltage-induced onset of positively
charged oxygen vacancies is hypothesized to be key to corrosion.  As will
be shown, and has been shown for the $\alpha$-Al$_2$O$_3$ model, the
oxygen vacancy orbital levels are filled at PZC conditions.  Hence it
is the vacancies with the highest-lying orbital levels that will be
most relevant to this study.  \color{black} We find that the
lowest onset potential of pitting\cite{pittingvoltage} occurs at potentials
sufficiently high that V$_{\rm O}^{2+}$ can exist in the outer region
of the oxide film, near the electrolyte.   This partially supports the
assumptions in the PDM that V${\rm O}^{2+}$ exists at the onset of pitting.

\color{black} This work continues in the vein of our previous work on explicit 
metal$|$passivating-layer interfaces for corrosion studies and for
batteries.\cite{corros1,pccp,selfdischarge}  The thickness of surface film
is explicitly treated as finite, and can be varied from the natural
passivating film thickness (no corrosion) to zero (inside corrosion pit),
as well as the intermediate, partially decomposed oxide thickness/regime, 
which may be particularly interesting for corrosion studies.
While the present work does not
involves cross-film electric fields and cross-film electron transfer
reaction,\cite{pccp,selfdischarge} the explicit metal$|$film models
are designed to permit future investigations in that direction. \color{black}

\section{Method and Models}

\subsection{DFT Details}
\label{dft}

Most DFT calculations in this work are conducted under ultra-high vacuum
(UHV) condition at T=0~K, using periodically replicated simulation cells
and the Vienna Atomic Simulation Package (VASP) version
5.4.\cite{vasp1,vasp1a,vasp2,vasp3} A 400~eV planewave energy cutoff and
a 10$^{-4}$~eV energy convergence criterion are enforced.  For simulation cells
associated with Fig.~\ref{fig1}c, {\it ab initio} molecular dynamics (AIMD)
simulations are conducted, using the same energy cutoff, a 10$^{-6}$~eV
energy convergence criterion, and a 0.5~fs time-step.  The temperature is
thermostat at T=400~K.  Most calculations apply the PBE functional.\cite{pbe}
In some cases, HSE06 is used for spot checks.\cite{hse06a,hse06b,hse06c}
\color{black} Screened long-range exchange DFT functionals have been shown to
work reasonably well for metals.\cite{marsman}
The slightly elevated temperature is needed because the PBE functional
yields overstructured water at T=300~K.\cite{dft_water}  Some dispersion force
corrections can also be used to correct for overstructuring, but those are
not applied here.  Spin-polarization is turned off because a net spin is
found to only accumulate in the metallic region, and it barely affects the
total energy.  \color{black}

The metal$|$oxide and oxide slab models are described in Table~\ref{table1}.
They contain vacuum regions in the $z$ direction, except the Fig.~\ref{fig1}c
configuration where the oxide slab is immersed in liquid water on both sides,
and a bulk oxide supercell.
When a Al(111) metal slab is present at the bottom of the cell, the bottom-most
layer of metal atoms is kept frozen.  The bulk and (110) surface structures
of $\gamma$-Al$_2$O$_3$ are taken from Ref.~\onlinecite{gamma}, except that
we enforce an orthonormal unit cell and set to zero the small deviation of
the $\gamma$ angle from 90$^o$.  The change in energy with this constraint
is less than 0.02~eV per $\gamma$-Al$_2$O$_3$ unit cell containing 40 atoms.
Apart from the lowest energy (110) surface cut from Ref.~\onlinecite{gamma}
(henceforth referred to as the ``A-cut,'') we also consider a slightly higher
surface energy cut of this surface (``B-cut,'' Fig.~\ref{fig2}).

We prepare the Al(111)$|$$\gamma$-Al$_2$O$_3$(110) interface by placing the
metal and the oxide slabs together, and then conducting AIMD simulations at
T=500~K for 28.2~ps using a 2~fs time step.  \color{black} The oxide lattice
constants are used, resulting in expanding the Al lattice constants
by 1\%. The oxide is 
approximately 18~\AA\, thick, assuming all atoms have 3.0~\AA\, radii.
As will be discussed in Fig.~\ref{fig2} below, this thickness is adequate
to yield a bulk-like band-gap region; however, an even thicker oxide will
be ideal in future studies. \color{black} The final AIMD configuration is 
quenched from 500~K to 50~K over 3.6~ps, and then optimized.  This annealing
procedure lowers the total energy of the 510-atom cell (Table~\ref{table1})
by 1.50~eV.  In the literature, a grid search approach, displacing the two
slabs laterally by a small increment followed by a configuration optimization,
and repeating using a different increment, has often been applied.
Thus we also attempt a restricted grid search by displacing the
initial oxide and metal later registry in 1~\AA\, increments in a 8$\times$8
grid.  Strictly speaking, the size of the interface cell (Table~1) should
require a 8$\times$24 grid, but the total energies are found to be so high,
and variations in work functions are so minor, that an expanded search is
deemed unlikely to produce an interfacial structure more favorable than the
AIMD-annealed one, or one that will yield a very different work function.
\color{black} In this work, we focus on a single, AIMD-equilibrated 
Al$|$$\gamma$-Al$_2$O$_3$ interface.  Unlike our previous work,\cite{corros1}
no model with both both metal$|$oxide and oxide$|$surface-hydroxide interfaces
is used, and the physics associated with the variation this interfacial
structure with properties like the oxide/water interface is not considered.
A more systematic study will be conducted in the future. \color{black}

Configurations associated with Fig.~\ref{fig1}c are initiated by freezing the 
$\gamma$-Al$_2$O$_3$ (110) slab at its DFT optimized configuration, with a
partially dissociated monolayer water from Ref.~\onlinecite{gamma} on top
and bottom, and applying grand canonical Monte Carlo (GCMC)\cite{towhee} with
a classical force fields\cite{clayff} to populate the vacuum region with liquid
water.  From the final GCMC configuration, we initiate AIMD simulations at
the slightly elevated temperature of T=400~K.  During AIMD trajectories,
H$^+$ transfer between H$_2$O and AlOH groups at oxide/water interfaces can
take place via the Grotthuss mechanism.  

Our simulations of oxide-coated Al metal in contact with liquid water are
relevant to both pitting corrosion of structural Al metal immersed in aqueous
media and to atmospheric corrosion, e.g., for microelectronics
applications.\cite{micro} At atmospheric conditions, the number of water layers
residing on the outer surface of the oxide film depends on the humidity but
usually exceed a few nanometer in thickness.\cite{atmos1,atmos2} We focus on
charge-neutral interfaces with no electric double layers in the electrolyte,
and assume that $\sim$12-18~\AA\, thick pure water films suffice.
A system size convergence check is conducted for the lower energy cut of
the (110) oxide surface.  Electrostatic potentials are calculated for
AIMD snapshots sampled every 100 time steps during a 40.0~ps trajectory
for the smaller, and during a 20.6~ps run for a larger cell with more water,
and are averaged.  \color{black} They are not computed every time step partly
due to computational cost, and partly because the electrostatic profile 
should not change significantly every 0.5~fs. \color{black} Atoms in the
AIMD simulation cells are not constrained; the spatial drifts of the oxide
slab during AIMD trajectories are negligable.  \color{black} Water dipole
orientation times can be significant on metal surfaces.\cite{gross}  
Our hydrophilic oxide surfaces, however, with its more well-defined
H-bond accepting and donating sites, likely form more well-defined hydrogen
bond networks with less global fluctuations, than pristine metal surfaces.
\color{black} 

Pure water$|$vacuum interfacial configurations for Fig.~\ref{fig1}d are taken
from Ref.~\onlinecite{surpot}.  In that work, we used atomic positions 
generated with classical force field, took 110 snapshots, and averaged
the electrostatic potential.  Pseudopotential core region shifts were applied
there to yield a zero potential outside noble atom cores,\cite{saunders}
although an ``uncorrected'' result of 2.80~V was also reported.  In this
work, we recompute the electrostatic profile without the corrections, using
the standard VASP procedure for calculating work functions (``LVHAR=.TRUE.'')
to ensure consistency and a proper cancellation of unwanted contributions
from the interface with vacuum.  We stress that \color{black} the value of 
\color{black} this water$|$vapor interfacial
contribution is strictly a DFT construct.\cite{surpot}  It cannot be
compared with the measured surface potential of water, \color{black} because DFT
calculations samples spatial regions inside the electron cloud of water
molecules, unlike electrochemical measurements which involve the electrostatic
potential experienced by ions that do not penetrate the interior of
water molecules.\cite{surpot,pnnl}  \color{black} Standard DFT work function
calculations in vacuum (Fig.~\ref{fig4}a contribution) also sample the interior
of atomic nuclei because they measure electronic, not ionic, properties.

Isolated, charge-neutral V$_{\rm O}$'s are introduced by removing one oxygen
atom each at various positions in a 3$\times$2$\times$2 bulk crystalline
oxide supercell.  There are several types of oxygen sites in
$\gamma$-Al$_2$O$_3$, with some coordinated to three Al$^{3+}$ ions and some to
four.  All 24 possible V$_{\rm O}$ defects in the primitive cell are considered.
We stress that the $\gamma$-Al$_2$O$_3$ model, and its oxygen vacancies,
are meant as model structures.  The true passivating oxides are typically
amorphous; until detailed experimental information about their structures
is available, our approach is to use model systems to elucidate 
voltage-function relations.  \color{black} For the same reason, we do
not focus on vacancy formation energies,\cite{freysoldt} or in general 
`structure-function relations,'' since the structures have not been
elucidated at the atomic length scale.  \color{black}

\begin{table}\centering
\begin{tabular}{l|l|r|l|r} \hline
system &  dimensions & stoichiometry & $k$-sampling & 
		Figure \\ \hline
Al(111)$|$$\gamma$(110) &  8.41$\times$24.23$\times$40.00 & 
	Al$_{294}$O$_{216}$ & 3$\times$1$\times$1  & Fig.~\ref{fig2}b \\
$\gamma$-Al$_2$O$_3$(110)$|$water &   16.82$\times$16.15$\times$32.00 & 
Al$_{128}$O$_{192}$(H$_2$O)$_{202}$ & 1$\times$1$\times$1  & Fig.~\ref{fig2}c \\
$\gamma$-Al$_2$O$_3$(110)$|$water &   16.82$\times$16.15$\times$38.00 & 
Al$_{128}$O$_{192}$(H$_2$O)$_{250}$ & 1$\times$1$\times$1  & Fig.~\ref{fig2}c \\
$\gamma$-Al$_2$O$_3$(110)$^*$$|$water &   16.82$\times$16.15$\times$35.00 & 
Al$_{128}$O$_{192}$(H$_2$O)$_{203}$ & 1$\times$1$\times$1  & Fig.~\ref{fig2}c \\
water$|$vacuum & 10.00$\times$10.00$\times$26.00 & 
(H$_2$O)$_{32}$ &  3$\times$1$\times$1  & Fig.~\ref{fig2}d \\
$\gamma$-Al$_2$O$_3$(110)$|$vacuum &   8.41$\times$24.23$\times$40.00 & 
Al$_{144}$O$_{216}$ & 3$\times$1$\times$1  & Fig.~\ref{fig2}e \\
$Al$(OH)$_3$$|$water &   17.86$\times$10.03$\times$38.00 & 
(Al(OH)$_3$)$_{48}$(H$_2$O)$_{145}$ & 1$\times$1$\times$1  & Fig.~\ref{fig7}b\\
$\gamma$-Al$_2$O$_3$ &   16.75$\times$16.82$\times$16.15 & 
Al$_{192}$O$_{288}$ & 1$\times$1$\times$1  & Fig.~\ref{fig8} \\
\hline
\end{tabular}
\caption[]
{\label{table1} \noindent
Computational details of the main simulation cells.  Dimensions are in
units of \AA$^3$. $^*$A higher surface energy cut of the slab (``B-cut'').
}
\end{table}

\subsection{Band Alignment and Voltage}
\label{voltage}

DFT is an electronic structure ground state theory and supports only one
$E_{\rm F}$.  The most readily available reference ``electrode'' is vacuum
(Fig.~\ref{fig1}a).  Following the Trasatti convention,\cite{trasatti} we
define the absolute voltage referenced against SHE (${\cal V}_e$) to be
\begin{equation}
{\cal V}_e = W/|e| - 4.44~V, \label{traseq}
\end{equation}
where $|e|$ is the electronic charge, $W$=$-E_{\rm F}$ is the work function,
and $E_{\rm F}$ is referenced to vacuum.  In general, the work function is
cumulatively modified by the electric double layer (EDL) and/or contact
potential at each interface,\cite{otani12} as well as the electric field
inside the electrolyte and electrode.  Under our flat band conditions, the
interfaces are uncharged and there is no electric field.  Assuming the oxide
and water regions are sufficiently thick, the electrostatic potential
contributions ($\phi$) should be additive:
\begin{equation}
\phi = (\phi_{\rm mf}+\phi_{\rm fv}) + \phi_{\rm fs} -\phi_{\rm fv}  
	+ \phi_{\rm sv} ,
\end{equation}
where ``m,'' ``f,'' ``s,'' and ``v'' stands for metal, (oxide) film, solvent,
and vacuum, respectively, and the four terms on the right correspond to
panels (b)-(e) in Fig.~\ref{fig1}.  Adding $-\phi_{\rm fv}$ cancels the
oxide$|$vacuum contribution that is present in Fig.~\ref{fig1}b which should
not exist in the true model system (Fig.~\ref{fig1}a).  Fig.~\ref{fig1}c
($\phi_{\rm fs}$) represents the only calculation without a vacuum region;
instead the plateau electrostatic reference value is extracted from the middle
of the liquid region.  

To our knowledge, only PZC's on pristine metal surfaces have been
computed\cite{juncheng} using AIMD methods developed for redox potential and
pK$_a$ calculaions.\cite{sulpizi,cheng2012,juncheng}  For example,
Ref.~\onlinecite{juncheng}
applies a ``computational hydrogen electrode'' (CHE) approach to calculate
the PZC of metals without surface oxides, instead of relying on the Trasatti
relation.  For our system, this approach would require both metal$|$oxide and
oxide$|$water interfaces in the same simulation cell, which is computationally
costly. This is one reason we have used an alternate method.  Note that, even
on pristine metal surfaces, implicit solvent models have been more frequently
applied to calculating PZC's.\cite{pristine,otani}  \color{black} With an
eye to future studies based on these models, we do not align the
metal$|$oxide interface using symmetric slab models without vacuum
regions.\cite{yoo2019}  The oxide$|$vacuum (or electrolyte) interface
plays significant roles in corrosion and related
disciplines,\cite{corros1,selfdischarge} and will be the focus of future
work based on our models.


\color{black}
The experimental approach used to determine the PZC is qualitatively different
from the DFT approach.  Bockris {\it et al.} modeled the passivating oxide
as a semiconductor.\cite{bockris}  The Fermi level was assumed to be pinned
at defect orbital levels, and the metal region was omitted.  A Mott-Schottky
approach was applied to vary the applied voltage to interpolate the potential
towards the vanishing ``depletion layer'' thickness condition associated with
the PZC.  One possible limitation of this approach is that, far away from the
PZC, the interpolation procedure assumed depletion layer widths that may
exceed the passivating oxide thickness.\cite{thickness} DFT calculations
typically apply simulation cells with crystalline oxide films with no or
few defects, unlike experimental oxide samples which exhibit a
range of defect level energies.  As a result, a metal ``electrode'' is
generally needed to determine the DFT Fermi level.\cite{costa,costa2,corros1}
\color{black}

\section{Results}

\subsection{Oxide$|$Vacuum Interface}



Fig.~\ref{fig2}a-b depicts the A- and B-cuts of $\gamma$-Al$_2$O$_3$ (110)
symmetric slabs, with the atomic coordinates relaxed to their lowest enegy
configurations.  The A-cut structure is consistent with
Ref.~\onlinecite{gamma} and gives a surface energy of 1.53~J/nm$^2$, within
0.01~J/nm$^2$ of the previous literature value.\cite{gamma}
The B-cut has a slightly higher surface energy of 1.76~J/nm$^2$.

\subsection{Al$|$Oxide Interface (Fig~\ref{fig1}b)}

\begin{figure}
\centerline{\hbox{ (a) \epsfxsize=2.00in \epsfbox{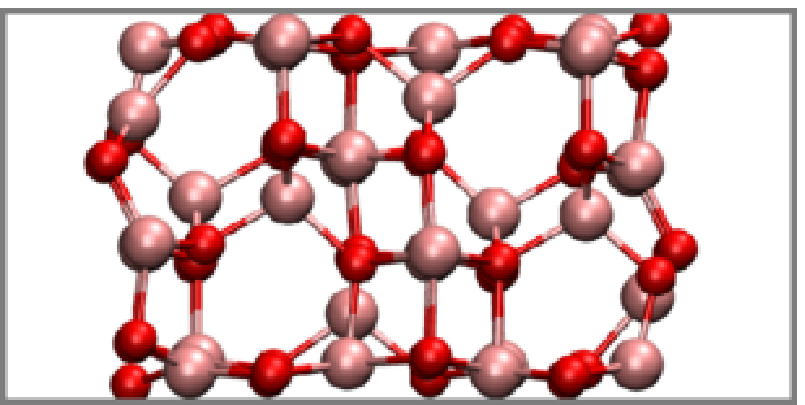} 
		       \epsfxsize=2.00in \epsfbox{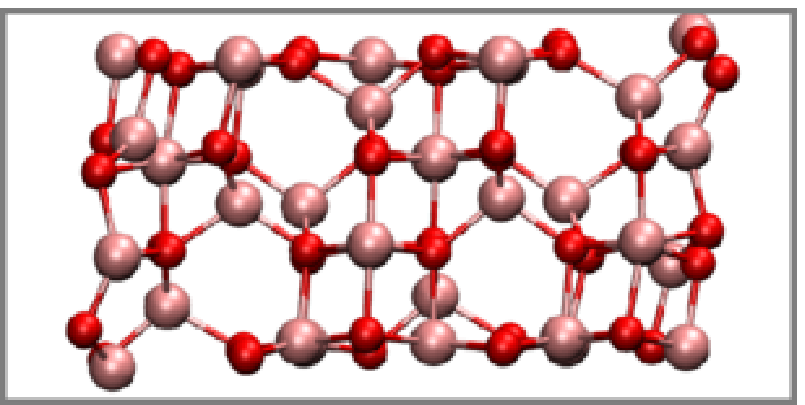} (b) }}
\centerline{\hbox{ (c) \epsfxsize=2.00in \epsfbox{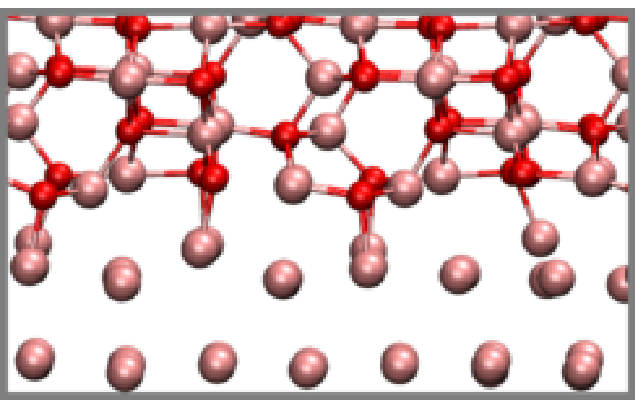} 
		       \epsfxsize=2.00in \epsfbox{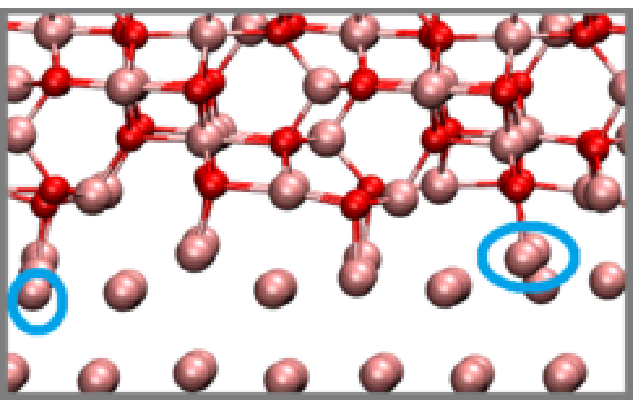} (d) }}
\centerline{\hbox{  \epsfxsize=4.00in \epsfbox{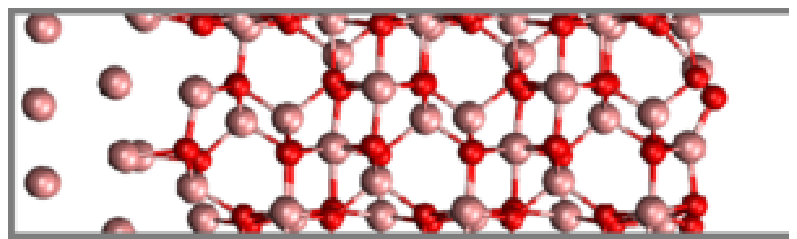} (e)}}
\centerline{\hbox{ \epsfxsize=4.50in \epsfbox{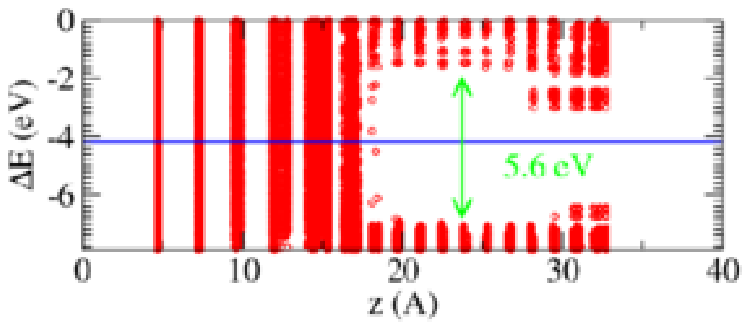} (f) }}
\caption[]
{\label{fig2} \noindent
(a) A-cut $\gamma$-Al$_2$O$_3$ (110); (b) B-cut $\gamma$-Al$_2$O$_3$ (110).
(c)-(d) A-cut on Al(111) with/without AIMD pre-equilibration.  The vacuum 
region atop the oxide layer is not shown.  There are two extra 
Al-O ionic bond in (c) than in (d), shown in blue circles.  (e) The entire
length of the AIMD-annealed A-cut slab.  (f) Local density of state
associated with panel (d).  The Al region ($z$$<$18~\AA) exhibits no
band gap while \color{black} the interior of \color{black} the oxide region
($z$$>$18~\AA) exhibits a $\sim$5.6~eV gap,
which is very similar to that computed for bulk $\gamma$-Al$_2$O$_3$
using the PBE functional.  Pink, red, and white spheres represent Al, O,
and H atoms, respectively.
}
\end{figure}

Fig.~\ref{fig2}c-d depict an A-cut directly optimized on the Al(111) surface,
and an A-cut oxide slab annealed with AIMD prior to optimization, respectively.
The entire length of the oxide-only simulation cell is depicted in
Fig.~\ref{fig2}e.  The AIMD-annealed metal$|$oxide interface configuration is
more favorable by 1.50~eV, due to the formation of two more Al-O ionic bonds
at the interface per simulation cell.  The work functions
are 4.05~eV and 4.18~eV before and after annealing, respectively.  This shows
that the AIMD-induced change in interfacial dipole surface density is small,
partly due to the fact that the increase in bonding is averaged over the
significant surface area of the simulation cell.

The bare Al(111) surface at this lattice constant has a work function of
3.97~eV when computed using the PBE functional.  Hence the contact potential,
which is the difference between work functions of the metal in contact/not
in contact with an oxide film without a permanent dipole moment\cite{pccp}
divided by $|e|$, is a modest +0.21~V for the AIMD-annealed (110) cut.  This
is small compared to other metal oxide interfaces examined in the DFT
literature.\cite{pccp,marks}
Our previous work on the $\alpha$-Al$_2$O$_3$(001)$|$vacuum interface predicts
a -1~eV contact potential;\cite{corros1} the larger value there likely 
arises from the more extensive metal-oxide contacts at that interface, which
has a higher surface atom density.

As discussed in the Method section, we have also performed a limited grid
search on the registry beween metal and oxide, without AIMD equilibration.
The total energies of these simulation cells are less favorable than the
AIMD-relaxed slab by 1.74$\pm0.06$~eV.  The most and least favorable
of these optimized configurations are 1.00 and 3.21~eV less favorable than
the AIMD-relaxed configuration.  The work functions range from 3.92 to 4.23~eV,
averaging to within 0.06$\pm$0.01~eV of the AIMD-relaxed slab value
(4.18~eV).  

Assuming that the overall variation of work functional with B-cut interfacial
structure is also small, we only conduct one work function calculation on
one B-cut structure without annealing.  We obtain a 4.02~eV work function,
which is only 0.16~eV lower than that for the the AIMD-relaxed A-cut
configuration.  This suggests that the work function, and voltage according
to Eq.~\ref{traseq}, \color{black} is not as strongly dependent on the details
of the metal$|$$\gamma$-Al$_2$O$_3$ and/or the $\gamma$-Al$_2$O$_3$$|$vacuum
interfaces, as in the case of $\alpha$-Al$_2$O$_3$.\cite{corros1} \color{black}

\subsection{Oxide$|$Water Interface (Fig.~\ref{fig1}c-d)}

\begin{figure}
\centerline{\hbox{ \epsfxsize=3.00in \epsfbox{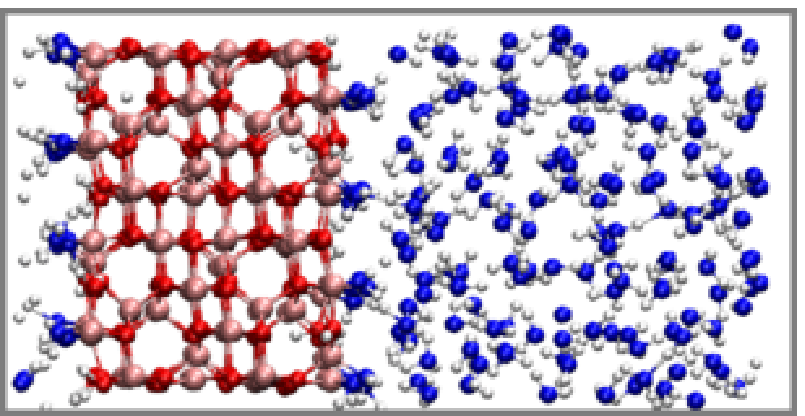} (a) }}
\centerline{\hbox{ \epsfxsize=3.00in \epsfbox{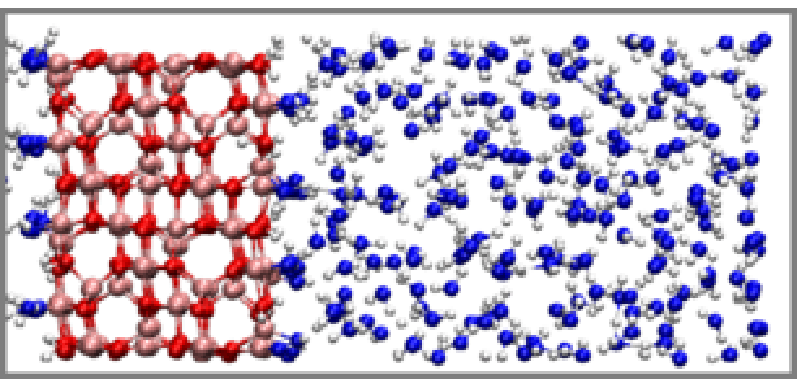} (b) }}
\centerline{\hbox{ \epsfxsize=3.00in \epsfbox{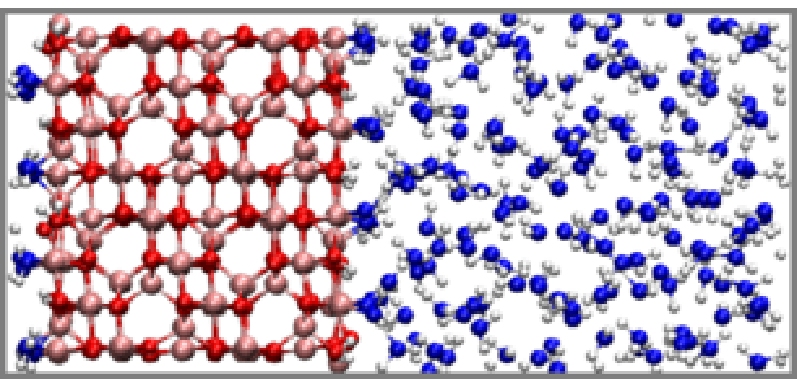} (c) }}
\caption[]
{\label{fig3} \noindent
Snapshot of the interface between water and (a) A-cut $\gamma$-Al$_2$O$_3$
(110); (b) same as (a) but with a thicker layer of water; (c) B-cut
$\gamma$-Al$_2$O$_3$ (110).  Pink, red, blue, and white spheres represent Al,
O, O(water), and H atoms, respectively
}
\end{figure}

\begin{figure}
\centerline{\hbox{ \epsfxsize=4.00in \epsfbox{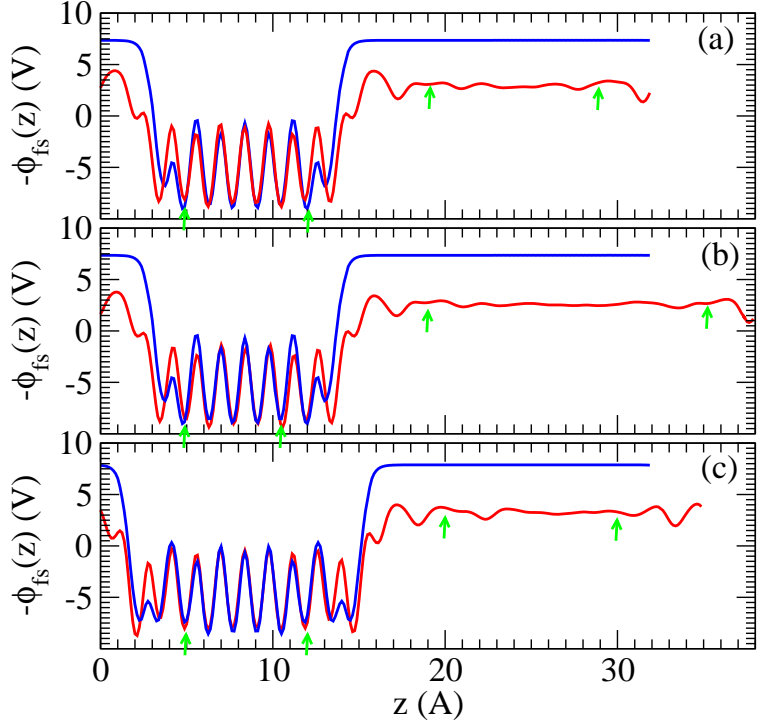} }}
\caption[]
{\label{fig4} \noindent
Electrostatic potential profiles ($\phi(z)$) between $\gamma$-Al$_2$O$_3$
(110) and water (red) or vacuum (blue).  (a) and (c) are for the
A- and B-cuts, respectively.  (b) is similar to (a) but has a thicker
water region.
}
\end{figure}

Fig.~\ref{fig3}a-c depict snapshots of AIMD simulations of the A- and
B-cuts of $\gamma$-Al$_2$O$_3$ immersed in water.  The corresponding
electrostatic potential profiles ($\phi(z)$), averaged in the $x$- and
$y$-directions, are depicted in Fig.~\ref{fig4}a-c, where results in
vacuum and in water are compared.  Only relative values in $\phi(z)$ are
significant; indeed the vacuum curves are shifted in voltage scale to match
the solid state region of the slab with water.

We integrate $-|e|\phi(z)$ in the oxide interior demarcated by the arrows
in Fig.~\ref{fig4}, and $-|e|\phi(z)$ in the water or vacuum plateau region
demarcated by the arrows there.  Then we subract the vacuum result from the
water value.  The shift in $-|e|\phi_{\rm fs}$ is -4.47~V for the smaller
A-cut model (Fig.~\ref{fig3}a,~\ref{fig4}a).  In the B-cut 
(Fig.~\ref{fig3}c,~\ref{fig4}c), they are -4.92~V.  $\phi(z)$ in the aqueous
region is not perfectly flat even when $z$ farthest from the solid surfaces.
This reflects a slightly incomplete sampling of the aqueous configurations of
the $\gamma$-Al$_2$O$_3$$|$water interfaces as well as the finite size of the
water region.  To examine finite size effects, we also consider an A-cut
simulation cell with a water film that is 6~\AA\, thicker (Fig.~\ref{fig3}b).
The shift between vacuum and water termination is -4.31~V (Fig.~\ref{fig4}b),
which is within 0.16~V of the system with a thinner water slab on this
surface (Fig.~\ref{fig4}a).  Henceforth we will report the 4.31~V associated
with the larger simulation cell; and treat the difference of 0.16~V between
the two cell sizes as a possible system size effect/systematic uncertainty in
our calculations.  $-|e|\Delta \phi_{fs}$=-4.31~eV and~-4.92~eV are added to
the work functions reported in the last section to obtain shifts arising
from the oxide/water (``{\it fs}'')
interfaces.  These ``work functions'' of the A- and B-cut oxide-covered
Al metal slabs become -0.13~eV and -0.70~eV, respectively.  However, the test
electron associated with these values are ejected into the aqueous region
(not vacuum). Therefore these values cannot be compared with experiments.

During AIMD simulations, we observe two OH groups point inwards, towards the
oxide interior, in the A-cut.   This is not observed in the monolayer water
configuration on this surface.\cite{gamma} However, these OH groups exhibit
no tendency of transferring their protons from the O-anions on the oxide
surface, which are coordinated to two Al$^{3+}$, to interior O-anions
coordinated to at least three Al$^{3+}$.

\subsection{Water$|$Vacuum Interface (Fig.~\ref{fig1}e)}

\begin{figure}
\centerline{\hbox{ \epsfxsize=4.00in \epsfbox{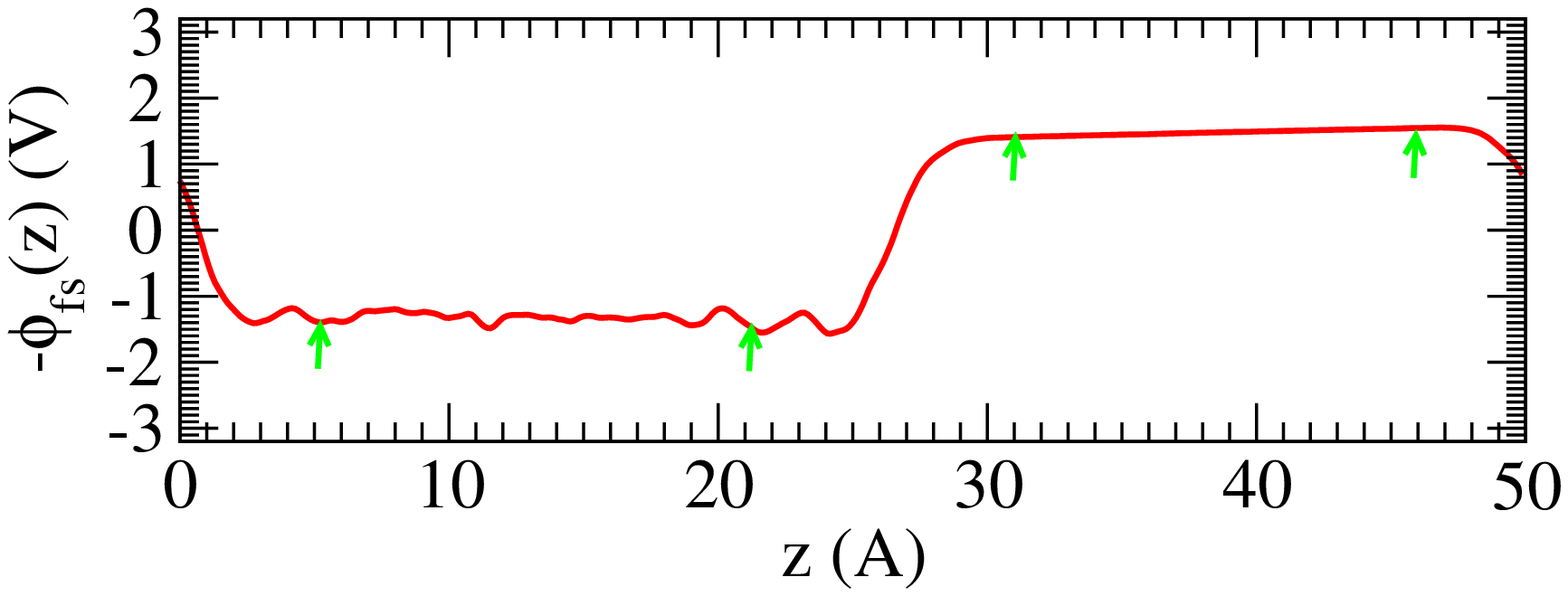} (a)}}
\centerline{\hbox{ \epsfxsize=3.80in \epsfbox{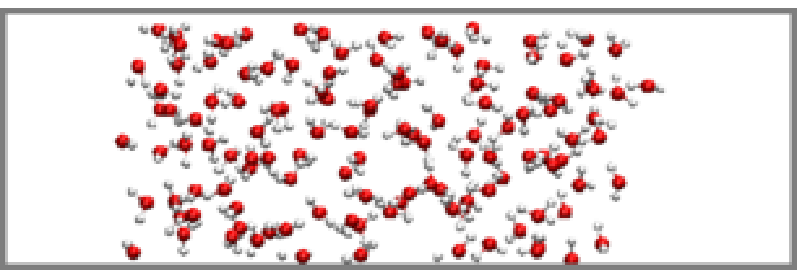} (b)}}
\caption[]
{\label{fig5} \noindent
(a) Electrostatic potential profiles ($\phi(z)$) at the water-vacuum interface.
(b) Snapshot of the water/vacuum interface.  Red and white spheres represent
O and H atoms, respectively.
}
\end{figure}

For this interface, we compute $\phi(z)$ by \color{black} averaging 
\color{black} the electrostatic profiles of a set of 110 classical force
field water configurations from previous calculations.\cite{surpot}
Fig.~\ref{fig5} depicts the average $\phi(z)$.  The small, 0.01~V/\AA\, slope
observed in the vacuum region is due to incomplete sampling, resulting in
a small net dipole moment in the water slab.  We take the value at the
center of the vacuum region as the plateau value to calculate the $\phi(x)$
shift between water and vacuum, yielding $\Delta \phi$=+2.80~V going from
liquid water to vacuum.  The water density in this model slab is actually
0.92~g/cm$^3$,\cite{surpot} not the canonical 1.0~g/cm$^{-3}$ value,
and the the potential difference should be proportional to the water density,
we further scale $\Delta \phi$ by 1/0.92, arriving at $\Delta \phi$=3.04~V.
\color{black} Note the classical force field MD trajectory used to generate the
configurations for DFT snapshot calculations use a simulation cell
with 12.5~\AA\, lateral dimensions (perpendicular to the interface).
The disperson force cutoff for the force field MD is less than half
the lateral box size, which results in the lower density. \color{black}

The significance of this +3.04~V shift is as follows.  There is no vacuum
region or water/vacuum interface in Fig.~\ref{fig1}c.  If there were,
the electrostatic potential would be shifted by
+3.04~V.  With this prediction, and using results from the last section,
Fig.~\ref{fig1}c, d,~and~e combine to give work functions of +2.91~eV
and 2.34~eV, respectively, for the water-covered A- and B-cuts.
In principal, these are measurable work functions, because the test
electrons are now ejected into the vacuum.

\subsection{Combining All Contributions}

\begin{figure}
\centerline{\hbox{ \epsfxsize=5.00in \epsfbox{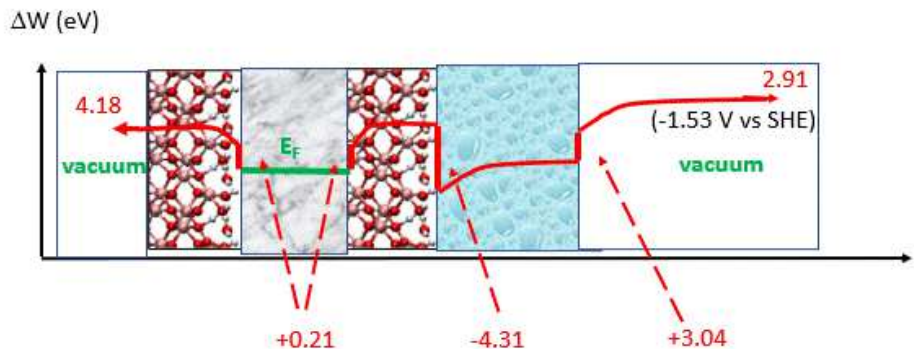}}}
\caption[]
{\label{fig6} \noindent
Schematic summary of the work function shifts at different interfaces, 
not drawn to scale.
}
\end{figure}

Converting the water-modified work functions to voltages using Eq.~\ref{traseq},
we obtain -1.53~V and -2.10~V vs.~SHE for the A- (Fig.~\ref{fig6}) and
B-cuts, respectively.  The former value, computed using the most stable
$\gamma$-Al$_2$O$_3$ (110) surface, is 0.52~V more negative than -1.01~V
reported in Ref.~\onlinecite{bockris}.  The other reported experimental
value (-0.7 vs.~SHE) is even smaller in magnitude.\cite{mccafferty99}
The B-cut value, computed using a higher (110) surface, is substantially
more negative still.

We predicted a -1.98~V vs.~SHE PZC value for $\alpha$-Al$_2$O$_3$(001)$|$vacuum
interface when a frozen monolayer of partially hydrolyzed H$_2$O exists on
that interface.\cite{corros1} However, as discussed earlier, the voltage
predicted there is too sensitive to the configuration of the frozen H$_2$O
molecules to be reliable.  In this work, the H$_2$O configurations are sampled
via AIMD simulations and should be correctly governed by the Boltzmann
distribution.  If we omit the liquid water contribution and only use the
single frozen layer of water of Ref.~\onlinecite{gamma}, the electrostatic
potential shift would only consist of the Fig.~\ref{fig1}b contribution, plus
the difference between Fig.~\ref{fig1}d with and without the frozen surface
water layer.  The work function shift due to adding the frozen water layer
is computed using a potential plot like Fig.~\ref{fig4}a (not shown).  It is
-1.13~eV. This correction reduces the work function of Fig.~\ref{fig1}b to
3.05~eV.  Using Eq.~\ref{traseq}, this translates into a PZC of -1.39~V, which
differ from the AIMD predictions by only +0.14~V.  This shows that, for this
$\gamma$-Al$_2$O$_3$ (110) surface, the predicted voltage does not
appear overly sensitive to the surface water content, unlike
$\alpha$-Al$_2$O$_3$ (111).

We propose that the discrepancy between the experimental and the DFT PZC is not
mainly due to uncertainties in the DFT calculations, or the fact that
the experimental work used the Mott-Schottky plot which requires extrapolation
from oxide film thicknesses that are beyond typical alumina film thicknesses
on Al metal.\cite{bockris}  Instead, we argue that discrepancy arises from
differences in the oxide stoichiometry.  The oxide films of
Ref.~\onlinecite{bockris}  has Al:O atomic ratios which vary with oxide depth.
The ratio should be 1:1.5 for ideal Al$_2$O$_3$, and close to the metal$|$oxide 
interface this ratio is indeed observed.  However, near the oxide$|$water
interface, this ratio can reach 1.8 to 2.0.\cite{bockris} This is rationalized
by the presence of protons, i.e., oxy-hydroxide content. Three H$^+$ need to
substitute for each Al$^{3+}$ to attain charge neutrality.  Another possibility
is that net charges exist inside the oxide film.\cite{bockris}  Our 
crystalline $\gamma$-Al$_2$O$_3$ oxide film model is not equipped to
deal with this latter possibility; we will defer investigation of space
charge effects and amorphous oxide models to future work.

\begin{figure}
\centerline{\hbox{ \epsfxsize=4.75in \epsfbox{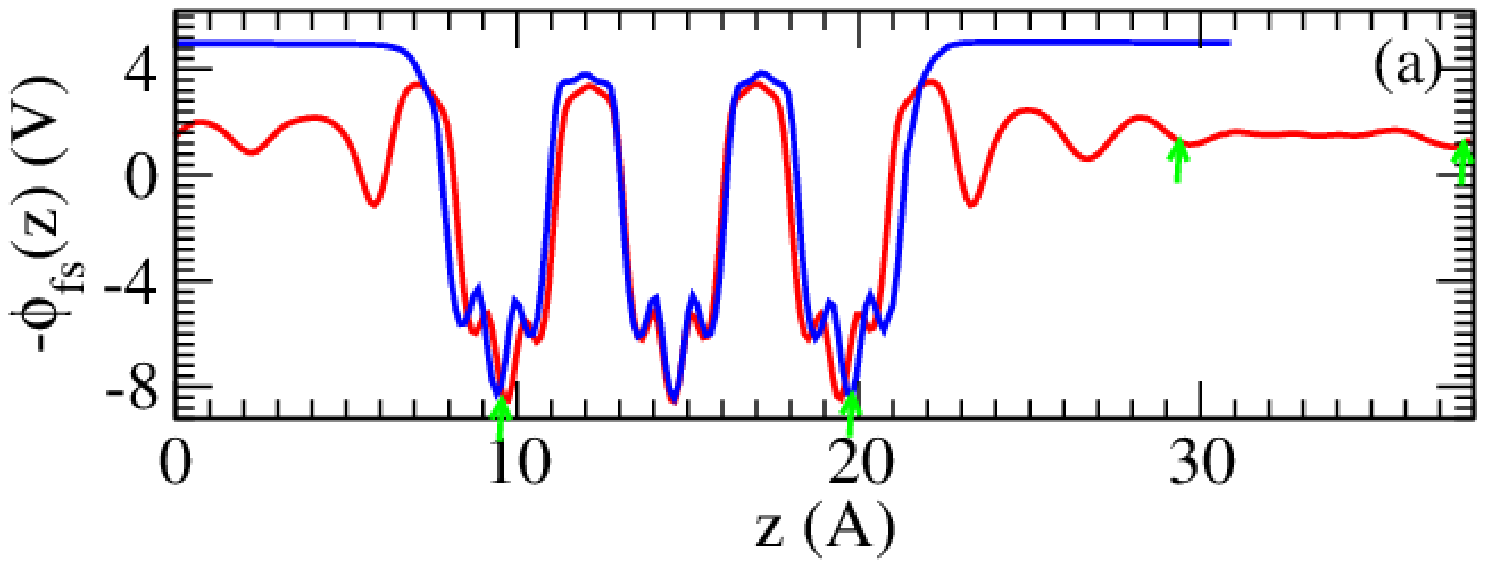} }}
\centerline{\hbox{ \epsfxsize=4.00in \epsfbox{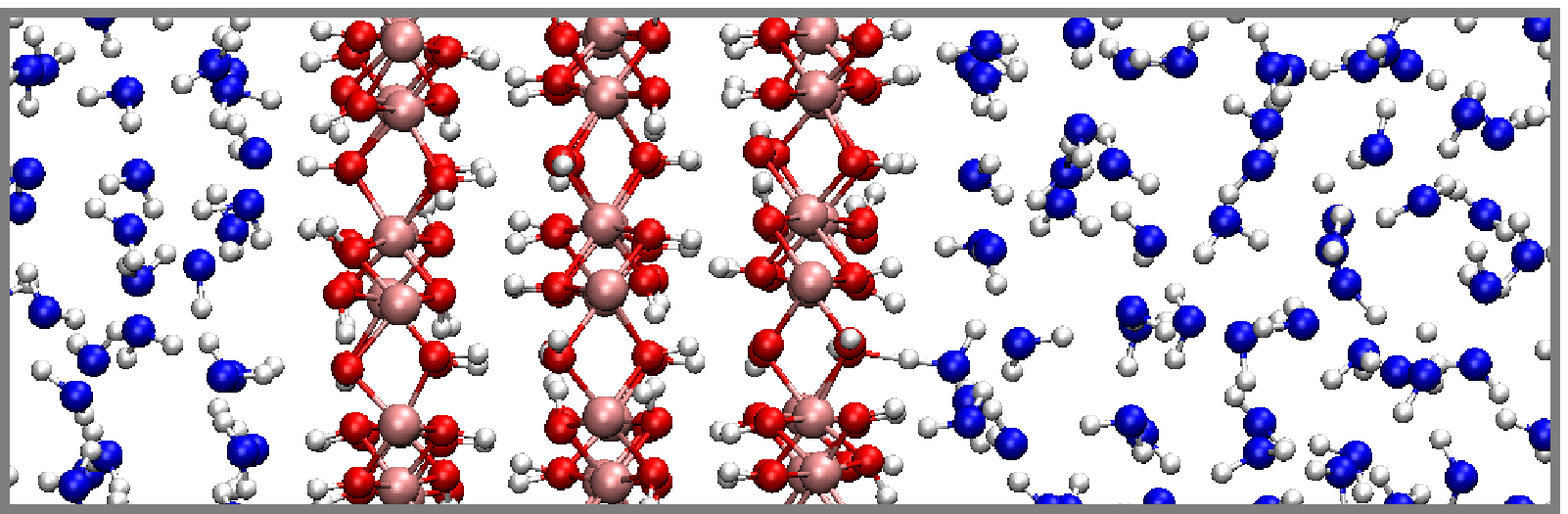} }}
\centerline{\hbox{ \hspace*{2in} (b) }}
\caption[]
{\label{fig7} \noindent
(a) Electrostatic potential profiles ($\phi(z)$) at the water-Al(OH)$_3$ (001)
interface.  Red and blue are with/without water, respectively.  (b) Snapshot
of the gibbsite (001)/water interfacial configuration.  Pink, red, blue, and
white spheres represent Al, O, O(water), and H atoms, respectively
}
\end{figure}

Two common forms of crystalline oxy-hydroxides are gibbsite (Al(OH)$_3$) and
boehmite (AlOOH).\cite{rosso}  Next we re-examine the difference between
the oxide$|$water (Fig.~\ref{fig1}c) and oxide$|$vacuum (Fig.~\ref{fig1}d)
contributions by substituting a Al(OH)$_3$ (001) slab\cite{rosso} for the
$\gamma$-Al$_2$O$_3$ slab.  This gibbsite model has the maximal hydrogen
content possible, and serves as a limiting case.
Fig.~\ref{fig7}a compares the electrostatic potentials computed for 
Al(OH)$_3$ (001) in the presence (snapshot shown in Fig.~\ref{fig7}b) and
absence of water, respectively.  The difference in the plateau region is
-3.18~V.  This is significantly less negative than the -4.31~V for the
$\gamma$-Al$_2$O$_3$ (110) A-cut surface.  If we simply apply -3.18~V to
the previous compilaton of interfacial contributions, we obtain a PZC value
of -0.40~V instead of -1.53~V.  This value is now substantially too
positive, rather than too negative, compared with experiments.  It shifts
the predicted PZC in the right direction, but overcorrects it.

This is a qualitative demonstration.  We have not attempted to calculate the
Fig.~\ref{fig1}b contribution with Al(OH)$_3$ (001) because Al(OH)$_3$ is
thermodynamically unstable with respect to Al metal and should react to
form H$_2$ gas.  Furthermore, the hydrogen content deduced in experiments
is not as high as in Al(OH)$_3$, which has a Al:O ratio of 1:3. 
Hence the difference between $\gamma$-Al$_2$O$_3$ and Al(OH)$_3$
is overestimated relative to experiments.  Nevertheless, Fig.~\ref{fig7} is
evocative and suggests that hydrogen content in the
film -- not just as hydroxyl groups on the surface layer but in subsurface
oxide layers as well -- can have significant effects on the PZC.  

Given the lack of definitive experimental oxide film structure and
stoichiometry at the atomic length scale, we propose the following
perspective.  Each DFT-based model with a specific passivating oxide
structure and interfaces should be considered a model system with
an intrinsic PZC.  The corrosion onset potential and other
corrosion-related properties should be referenced to this PZC
rather than the experimental value.  For better comparison with experiments,
we should devise atomic lengh-scale models that yield the experimental PZC.

\subsection{Oxygen Vacancy Orbital Level}

Finally, we elucidate the conditions under which oxygen vacancies may have net
positive charges in our model (Fig.~\ref{fig1}f).  Such charged vacancies are
foundational within the PDM corrosion theory.\cite{macdonald81a} As in
Ref.~\onlinecite{corros1}, we define $E_{\rm VoVBE}$ to be the V$_{\rm O}$
energy level above the oxide valence band edge (VBE), and $E_{\rm FVBE}$ as
the difference between the VBE energy level and the Fermi level ($E_{\rm F}$).
The former is computed in simulation cells without Al metal, while the
latter is only relevant for simulation cells with metal electrodes.  From
Fig.~\ref{fig2}f, the local density of state (LDOS) for the configuration of
Fig.~\ref{fig2}d, we find that $E_{\rm FVBE}$=2.89~eV in the ``bulk-like''
region of the oxide in Fig.~\ref{fig2}f (20~\AA$<$$z$$<$29~\AA), but is
2.26~eV near the oxide outer surface ($z$$>$29~\AA).
\color{black} The LDOS is constructed by decomposition both occupied and
unoccupied orbitals on to atomic contributions.\cite{solid} \color{black}

\begin{figure}
\centerline{\hbox{ \epsfxsize=4.00in \epsfbox{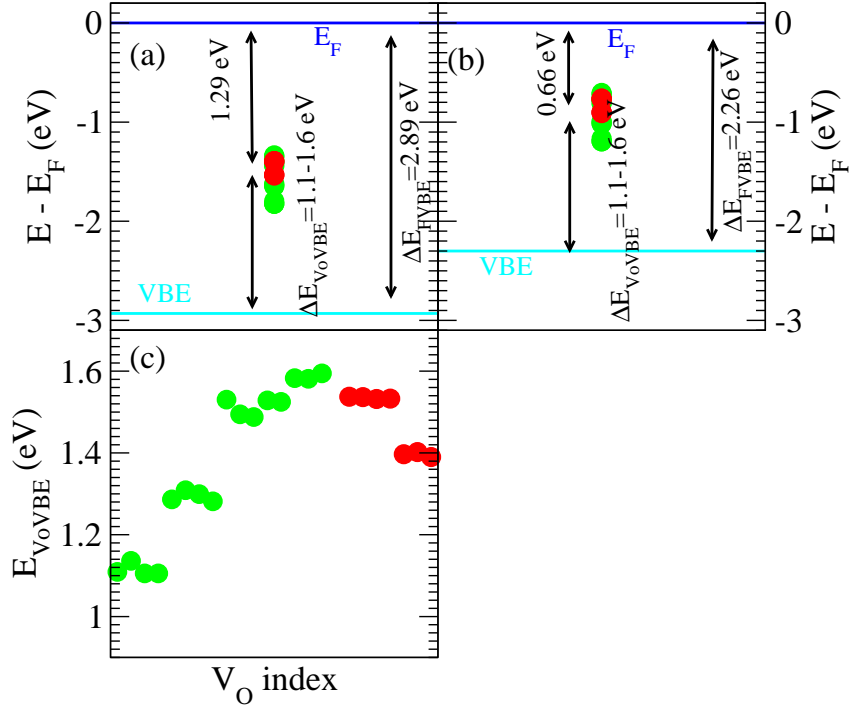} }}
\caption[]
{\label{fig8} \noindent
(a)-(b) Schematic aligning the charge-neutral V$_{\rm O}$ orbital levels
to the metal $E_{\rm F}$, for the bulk-like and surface oxide regions
of Fig.~\ref{fig2}f), respectively.  
(c) Orbital levels for the 24 distinct optimized vacancies, relative to 
the VBE.  Red and green refer to 3- and 4-coordinated O-sites, respectively.
}
\end{figure}

Fig.~\ref{fig8}c depicts $E_{\rm VoVBE}$ for the 24 possible distinct 
V$_{\rm O}$ in the bulk $\gamma$-Al$_2$O$_3$ supercell (Table~\ref{table1}).
Unlike $\alpha$-Al$_2$O$_3$, not all O-sites are equilivalent; some O~atoms in
the crystal are 3-coordinated while others are 4-coordinated.  The defect
orbital levels reside in 4~bands centering around 1.1, 1.3, 1.4, and 1.6~eV
above the VBE. The highest-lying defect orbitals correspond to 4-coordinated
O-sites.  The V$_{\rm O}$ formation energies of these defects (not orbital
levels) span a range of 0.28~eV.  For our calculations, we pick the highest
lying defect orbital level so that, if this V$_{\rm O}$ is occupied by
electrons and charge-neutral, all other V$_{\rm O}$ choices are also
charge-neutral.  This yields $E_{\rm VoVBE}$=1.60~eV.

To translate the $E_{\rm VoVBE}$ predictions to the $E_{\rm FVBE}$ 
associated with the metal/oxide/water PZC configuration (Fig.~\ref{fig1}a
via Fig.~\ref{fig1}b-d), we subtract the 1.60~eV relevant to the highest
charge-neutral vacancy orbital energy level from $E_{\rm FVBE}$, yielding
1.29~eV in the oxide interior and 0.66~eV near the oxide outer surface.  This
means that all V$_{\rm O}$ are at least 1.29~eV/0.66~eV below the
Fermi level at PZC conditions (Fig.~\ref{fig8}a-b); they all contain defect
orbitals occupied by electrons and are charge-neutral.  This assessment
qualitatively agrees with $\alpha$-Al$_2$O$_3$, where the V$_{\rm O}$ level
is 0.60~eV below the Fermi level.\cite{corros1}

The PZC is estimated above to be -1.53~V vs.~SHE.  To push O-vacancy orbitals
in the oxide interior above the Fermi level, so that they acquire positive
charges, requires raising the voltage by at least 1.29~V, to -0.24~V vs.~SHE.
The -0.24~V value for this model similar to the measured onset potential of
pitting in Al metals of $\sim$-0.3~V,\cite{bockris}, although it is higher
than the minimum pitting potential of -0.50~V reported
elsewhere.\cite{pittingvoltage} At or above -0.24~V, our model system would
exhibit V$_{\rm O}^{2+}$.  If we had used the experimental PZC of $\sim$-1.0~V,
a value of +0.29~V would be needed to create V$_{\rm O}^{2+}$, which is
significantly above the experimental pitting potential of -0.3~V.  As
discussed above, for internal consistency, we should use the computed PZC
to characterize oxide film properties like $V_{\rm O}$ orbital levels in
DFT models.  For O vacancies oxide near the oxide outer surface in
Fig.~\ref{fig2}f, V$_{\rm O}^{2+}$ can occur at much lower potentials of
(-1.53+0.66)=-0.87~V.

This above estimates are derived using the flat-band approximation, where the
oxide film electrostatic potential is raised uniformly with increasing
applied potential.  In reality, when a charged surface (e.g., enhanced
ion concentrations in EDL in the liquid electrolyte) exists to
raise the overall voltage, the oxide region near the liquid electrolyte is at
higher energy levels than the region near the metal surface due to
the electric field across the oxide film.  Hence the outer
region of the passivating oxide may begin to exhibit oxygen vacancies that
are positively charged at lower potentials than those near the metal surface.

Finally, we address DFT functional accuracies.  The more accurate HSE06
functional\cite{hse06a,hse06b,hse06c} was previously applied to spot-check
PBE predictions for the Al(111)$|$$\alpha$-Al$_2$O$_3$ (111)
model.\cite{corros1}  Compared to PBE calculations, the V$_{\rm O}$ orbital
level was found to shift up into the band gap by $\Delta E_{\rm VoVBE}$=0.78~eV.
For the bulk $\gamma$-Al$_2$O$_3$ 3$\times$2$\times$2 supercell, we find that
HSE06 predicts $E_{\rm VoVBE}$=2.25~eV at one of the V$_{\rm O}$ site where
PBE predicts $E_{\rm VoVBE}$=1.60~eV.  Thus $\Delta E_{\rm VoVBE}$ is 0.65~eV,
which is reasonably similar to the 0.78~eV value predicted for
$\alpha$-Al$_2$O$_3$.

HSE06 also shifted the VBE edge downwards relative to $E_{\rm F}$, i.e., it
increased $\Delta E_{\rm  FVBE}$ for Al(111)$|$$\alpha$-Al$_2$O$_3$ (111), by
1.14~eV.  The net effect was a downward revision of the occupied V$_{\rm O}$
vacancy level by 0.36~eV.  Unfortunately, the Al(111)$|$$\gamma$-Al$_2$O$_3$
(110) interface model simulation cell (Table~\ref{table1}), which is a focus
of the present work, is too large to permit a HSE06 spot check of
$E_{\rm FVBE}$.  

If we assume the overall magnitude of the HSE06 correction to PBE for
$\alpha$-Al$_2$O$_3$ also corrects for the current $\gamma$-Al$_2$O$_3$
system, it would further lower the V$_{\rm O}$ orbital energy by 0.36~V 
relative to E$_{\rm F}$, compared to PBE values.  With this correction,
it would require an extra +0.36~V, or a total of -0.51~V instead of -0.87~V,
to raise the V$_{\rm O}$ level above E$_{\rm F}$ in the outer oxide
region.  While this value is still within the range of the -0.5~V pitting
onset potential for aluminum,\cite{pittingvoltage} the interior of the
oxide would require at least (-0.24+0.36)=+0.12~V to acquire positive
charges.  This semi-quantitative discussion of DFT accuracy appears to 
reinforce our conclusion that the outer regions of the oxide is more
likely to support V$_{\rm O}^{2+}$ vacancies.  However, as the oxide starts
to break down from its outer, electrolyte-facing surface, the interior
regions should become more ``surface-like'' and become more susceptable
to creation of V$_{\rm O}^{2+}$.

\section{Conclusion}
\label{Conclusion}

In this work, we have applied DFT/PBE methods to calculate the 
potential-of-zero-charge (PZC) of a model system relevant to aluminum
corrosion.  Our models consist of Al(111) slabs covered with thin, low
energy cuts of $\gamma$-Al$_2$O$_3$ (110) films immersed in liquid water.
They mimic Al metal passivated with amorphous oxide films in contact
with multiple layers of water, which are relevant to both inundated and
high-humidity atmospheric corrosion conditions.  The models have no explicit
surface charge either on the metal or the oxide film surface and are
therefore simultaneously at the PZC of the electrode and the pH-of-zero-charge
of the oxide.  By splitting this system with two interfaces into multiple
binary systems with one interface each, and summing the contributions, we
predict that the PZC resides at -1.53~V vs.~SHE.  The systematic error may
be on the order of 0.16~V and is mostly attributed to AIMD simulations of
liquid water at Al$_2$O$_3$ surfaces.  The prediction is more negative than
the $\sim$-0.7 to~-1.0~V experimental values reported in the
literature.\cite{bockris,mccafferty99}  Using a higher energy cut
of the (110) surface slightly alters the DFT predicted PZC value
to an even more negative -2.10~V, which is a measure of the variation
of PZC with interfacial structure.  We argue that the predicted PZC value
is too negative because the DFT oxide model lacks sufficient proton
content.\cite{bockris}  Adopting the other extreme, namely a gibbsite surface
film (Al(OH)$_3$) with too much hydrogen content compared to measurements,
indeed yields a significantly more positive PZC of -0.49~V.  

Under our PZC condition, the highest lying oxygen vacancy (V$_{\rm O}$)
orbital level resides at 1.29 to 0.66~eV below the Fermi level, depending on
whether the vacancy is in the oxide interior or near the oxide$|$electrolyte
interface.  Within a flat-band approximation, this implies that
raising the potential above -0.24~V or~-0.87~V vs.~SHE is sufficient to raise
the V$_{\rm O}$ orbitals above the Fermi level and render at least some oxygen
vacancies positively charged.  V$_{\rm O}^{2+}$ are integral parts of the
point defect model, one of the theories  widely used to analyze pitting
corrosion kinetics.\cite{macdonald81a} Our predictions for oxygen vacancy
near the oxide outer surface are consistent with the PDM assumption that
the reported minimal onset potential for pitting for Al of
-0.5~V vs.~SHE,\cite{pittingvoltage} which is higher than the -0.87~V
needed to create V$_{\rm O}^{2+}$ there.  We argue that, as the passivation
oxide starts to break down, more of the interior oxide becomes surface-like
and becomes more susceptable to V$_{\rm O}^{2+}$ formation.  Future
work along these lines will involves inclusion of an explicit electric
field across the oxide, which enhances V$_{\rm O}^{2+}$ creation
in the oxide film near the liquid electrolyte but less so near the
metal surface; and the consideration of true amorphous oxide models, 
which may yield a wider distribution of V$_{\rm O}$ orbital levels
than the $\gamma$-Al$_2$O$_3$ model applied in this work, as well as
a non-zero space charge inside the oxide.  Computationally
speaking, we find that each aluminum metal/passivating oxide DFT model
exhibits a different PZC.  In the future, we should construct interfacial
models with predicted PZC's that agree with the experimental value.  In
other words, the computed PZC should be used as a constraint on the model,
not treated as a prediction.  Finally, this work also provides a benchmark for
future implicit solvent calculations.\cite{campbell}

\section*{Acknowledgement}

We thank Anastasia Ilgen, Jacob Harvey, and Quinn Campbell for useful
suggestions.  This work is funded by the Advanced Strategic Computing (ASC)
Program.  Sandia National Laboratories is a multi-mission laboratory managed
and operated by National Technology and Engineering Solutions of Sandia, LLC,
a wholly owned subsidiary of Honeywell International, Inc., for the U.S.
Department of Energy’s National Nuclear Security Administration under contract
DE-NA0003525.  This paper describes objective technical results and analysis.
Any subjective views or opinions that might be expressed in the document do
not necessarily represent the views of the U.S. Department of Energy or the
United States Government.




\begin{references}

\bibitem{corrbook}
R.W.~Revie (eds) 
{\it Uhlig's Corrosion Handbook}, Third Edition.  (Wiley, 2011)

\bibitem{mccafferty10}
E.~McCafferty, {\it Electrochim.~Acta} {\bf 55}, 1630-1637 (2010).

\bibitem{mccafferty99}
E.~McCafferty, {\it J.~Electrochem.~Soc.} {\bf 146}, 2863 (1999).

\bibitem{natishan88}
P.M.~Natishan, E.~McCafferty, and G.K.~Hubler,
{\it J.~Electrochem.~Soc.} {\bf 135}, 321 (1988)

\bibitem{macdonald81a}
C.Y.~Chao, L.F.~Lin, and D.D. Macdonald, {\it J.~Electrochem.~Soc.} 
{\bf 128}, 1187 (1981).

\bibitem{corros1}
K.~Leung, {\it J.~Electrochem.~Soc.} {\bf 168}, 031511 (2021).

\bibitem{pristine}
K.~Letchworth-Weaver and T.A.~Arias, {\it Phys.~Rev.~B}, {\bf 86}, 075140
(2012).

\bibitem{juncheng}
X.-Y. Li, A.~Chen, X.-H.~Yang, J.-X.~Zhu, J.-B.~Le, and J.~Cheng, 
{\it J.~Phys.~Chem.~Lett.} {\bf 12}, 7299 (2021).

\bibitem{taylor}
C.D.~Taylor, M.J.~Janik, M.~Neurock, and R.G.~Kelly,
{\it Mol.~Sim.} {\bf 33}, 429 (2007).

\bibitem{otani}
K.~Kano, S.~Hagiwara, T.~Igarashi, and M.~Otani, {\it Electrochim.~Acta}
{\bf 377}, 138121 (2021).

\bibitem{bockris}
J.O'M.~Bockris, and Y.K.~Kang, {\it J.~Solid State Electrochem.} {\bf 1}, 17
(1997).

\bibitem{campbell}
Q.~Campbell and I.~Dabo, {\it Phys.~Rev.~B}, {\bf 96}, 205308 (2017).

\bibitem{pittingvoltage}
J.B.~Bessone, D.R.~Salinas, C.E.~Mayer, M.~Ebert \& W.J.~Lorenz,
{\it Electrochim.~Acta} {\bf 37}, 2283 (1992).

\bibitem{marks}
X.-X.~Yu, and L.D.~Marks, {\it Corrosion}, {\bf 75}, 152-166 (2019).

\bibitem{costa}
P.~Cornette, D.~Costa, and P.~Marcus,
{\it J.~Electrochem.~Soc.} {\bf 167}, 161501 (2020).

\bibitem{costa2}
D.~Costa, T.~Ribeiro, P.~Cornette, and P.~Marcus,
{\it J.~Phys.~Chem.~C}, {\bf 120}, 28607 (2016).

\bibitem{costa3}
D.~Costa, T.~Ribeiro, F.~Mercuri, G.~Pacchioni, and P.~Marcus,
{\it Adv.~Mater.~Interfaces} {\bf 1}, 1300072 (2014).

\bibitem{liu2021}
M.~Li, Y.~Jin, B.~Chen, C.~Leygraf, L.~Wang, and J.~Pan, 
{\it J.~Electrochem.~Soc.} {\bf 168}, 081508 (2021).

\bibitem{kadowaki}
M.~Kadowaki, A.~Saengdeejing, I.~Muto, Y.~Chen, T.~Doi, K.~Kawano, Y.~Sugawara,
and  N.~Hara, {\it J.~Electrochem.~Soc.} {\bf 168}, 111503 (2021).

\bibitem{macdonald81b}
L.F.~Lin, C.Y.~Chao, and D.D. Macdonald, {\it J.~Electrochem.~Soc.} 
{\bf 128}, 1194 (1981).

\bibitem{macdonald16a}
D.D.~Macdonald and X.~Lei.
{\it J.~Electrochem.~Soc.}, {\bf 163} C738 (2016).

\bibitem{macdonald15}
S.~Sharifi-Asl, F.~Mao, P.~Lu, B.~Kursten, and D.D.~Macdonald.
{\it Corrosion Sci.} {\bf 98}, 708 (2015).

\bibitem{macdonald16b}
P.~Lu, R.~Engelhardt, B.~Kursten, and D.D.~Macdonald.
{\it J.~Electrochem.~Soc.} {\bf 163} C156 (2016).

\bibitem{theory1}
G.S.~Frankel, 
{\it J.~Electrochem.~Soc.} {\bf 145}, 2186 (1998).

\bibitem{theory2}
M.~Bojinov,G.~Fabricius, T.~Laitinen, K.~Makela, T.~Saario, and G.~Sundholm,
{\it Electrochim.~Acta} {\bf 45}, 2029 (2000).

\bibitem{theory3}
A.~Seyeux, V.~Maurice, and P.~Marcus, 
{\it J.~Electrom.~Soc.} {\bf 160}, C189 (2013).

\bibitem{theory4}
A.~Couet, A.T.~Motta, and A.~Ambard, {\it Corros.~Sci.} {\bf 100}, 73 (2015).

\bibitem{theory5}
A.~Momeni and J.C.~Wren, {\it Faraday Discuss.} {\bf 180}, 113 (2015).

\bibitem{theory6}
C.~Batallion, F.~Bouchon, C.~Chainais-Hillairet, C.~Desgranges, E.~Hoarau, 
F.~Martin, S.~Perrin, M.~Tupin, and J.~Talandier, {\it Electrochim.~Acta}
{\bf 55}, 4451 (2010).

\bibitem{theory7}
E.~McCafferty, {\it Corros.~Sci.} {\bf 45}, 1421 (2003).

\bibitem{otani12}
N.~Bonnet, T.~Morishita, O.~Sugino, and M.~Otani,
{\it Phys.~Rev.~Lett.}, {\bf 109}, 266101 (2012).

\bibitem{pccp}
K.~Leung,
{\it Phys.~Chem.~Chem.~Phys.} {\bf 22}, 10412 (2020).

\bibitem{selfdischarge}
K.~Leung, L.C.~Merrill, and K.L.~Harrison, 
{\it J.~Phys.~Chem.~C} {\bf 126}, 8565 (2022).

\bibitem{shluger1}
O.A.~Dicks, J.~Cottom, A.L.~Shluger, and V.V.~Afanas'ev.
{\it Nanotechnology} {\bf 30}, 205201 (2019).

\bibitem{shluger2}
J.~Strand, M.~Kaviani, D.~Gao, A.~El-Sayed, V.V. Afanas'ev, and A.L.~Shluger.
{\it J.~Phys.: Condens.~Matter}, {\bf 30}, 233001 (2018).

\bibitem{persson}
M.~Aykol and K.A.~Persson,
{\it ACS Appl.~Mater.~Interfaces} {\bf 10}, 3039 (2018).

\bibitem{thickness}
J.~Evertsson, F.~Bertram, F.~Zhang, L.~Rullik, L.R.~Merte, M.~Shipilin,
M.~Soldemo, S.~Ahmadi, N.~Vinogradov, F.~Carla, J.~Weissenrieder,
M.~G\"{o}thelid, J.~Pan, A.~Mikkelsen, J.-O.~Nilsson, and E.~Lundgren,
{\it Appl.~Sur.~Sci.} {\bf 349}, 826 (2015).

\bibitem{vasp1}
G.~Kresse and J.~Furthm\"{u}ller, {\it Phys.~Rev.~B}, {\bf 54}, 11169 (1996).

\bibitem{vasp1a}
G.~Kresse and J.~Furthm\"{u}ller, {\it Comput.~Mater.~Sci.}, {\bf 6},
15-50 (1996).

\bibitem{vasp2}
G.~Kresse and D.~Joubert, {\it Phys.~Rev.~B}, {\bf 59}, 1758 (1999).

\bibitem{vasp3}
J.~Paier, M.~Marsman, and G.~Kresse, {\it J.~Chem.~Phys.}, {\bf 127},
024103 (2007).

\bibitem{pbe}
J.P.~Perdew, K.~Burke, and M.~Ernzerhof, {\it Phys. Rev. Lett.}, 
{\bf 77}, 3865 (1996).

\bibitem{hse06a}
J.~Heyd, G.E.~Scuseria, and M.~Ernzerhof,
{\it J.~Chem.~Phys.}, {\bf 118}, 8207 (2003).

\bibitem{hse06b}
J.~Heyd, G.E.~Scuseria, and M.~Ernzerhof,
{\it J.~Chem.~Phys.}, {\bf 124}, 219906 (2006).

\bibitem{hse06c}
O.A.~Vydrov, J.~Heyd, A.V.~Krukau, and G.E.~Scuseria,
{\it J.~Chem.~Phys.}, {\bf 125}, 074106 (2006).

\bibitem{marsman}
M.~Marsman, J.~Paier, A.~Stroppa, and G.~Kresse,
{\it J.~Phys.~Condens.~Matter}, {\bf 20}, 064201 (2008).

\bibitem{dft_water}
See, e.g., E.~Schwegler, J.C.~Grossman, F.~Gygi, and G.~Galli,
{\it J. Chem. Phys.}, {\bf 121}, 5400 (2004).

\bibitem{gamma}
B.F.~Ngouana-Wakou, P.~Cornette, M.C.~Valero, D.~Costa,
and P.~Raybaud, {\it J.~Phys.~Chem.~C} {\bf 121}, 10351 (2017).

\bibitem{towhee}
M.G.~Martin and A.P.~Thompson, A.P.  {\it Fluid Phase Equil.}
{\bf 217}, 105 (2004).

\bibitem{clayff}
R.T.~Cygan, J.-J.~Liang, and A.G.,~Kalinichev {\it J.~Phys.~Chem.~B} {\bf 108},
1255, (2004).

\bibitem{micro}
Y.Y.~Cheng, S.X.~Lu, W.G.~Xu, and H.D.~Wen,
{\it RSC Adv.} {\bf 5}, 15387 (2015).

\bibitem{atmos1}
R.M.~Katona, S.~Tokuda, J.~Perry, and R.G.~Kelly,
{\it Corrosion Sci.} {\bf 175}, 108849 (2020).

\bibitem{atmos2}
R.M.~Katona, A.W.~Knight, E.J.~Schindelholz, C.R.~Bryan, R.F.~Schaller,
and R.G.~Kelly, {\it Electrochim. Acta} {\bf 370}, 137696 (2021).

\bibitem{gross}
S.~Sakong and A.~Gross, {\it Phys.~Chem.~Chem.~Phys.} {\bf 22}, 10431 (2020).

\bibitem{surpot}
K. Leung, {\it J.~Phys.~Chem.~Lett.} {\bf 1}, 496 (2010).

\bibitem{freysoldt}
C.~Freysoldt, B.~Grabowski, T.~Hickel, J.~Neugebauer, G.~Kresse, 
A.~Janotti, and C.G.~Van de Walle, {\it Rev.~Mod.~Phys.} {\bf 86}, 253 (2014).

\bibitem{saunders}
V.R.~Saunders, C.~Freyria-Fava, R.~Dovesi, L.~Salasco, C.~Roetti, 
{\it Mol. Phys.} {\bf 77}, 629 (1992).

\bibitem{pnnl}
S.M.~Kathmann, I.F.W.~Kuo, C.J.~Mundy, and G.K.~Schenter, 
{\it J.~Phys.~Chem.~B} {\bf 115}, 4369 (2011).

\bibitem{trasatti}
S.~Trasatti, {\it J.~Electroanal.~Chem.} {\bf 209}, 417 (1986).

\bibitem{sulpizi}
J.~Cheng, X.~Liu, J.~VandeVondele, M.~Sulpizi, and M.~Sprik. 
{\it Acc. Chem. Res.}, {\bf 47}, 3522 (2014).

\bibitem{yoo2019}
S.-H.~Yoo, N.~Siemer, M.~Todorova, D.~Marx, and J.~Neugebauer, 
{\it J.~Phys.~Chem.~C} {\bf 123}, 5495 (2019).

\bibitem{cheng2012}
J.~Cheng and M.~Sprik, 
{\it Phys.~Chem.~Chem.~Phys.} {\bf 14}, 11245 (2012).


\bibitem{rosso}
M.~Sassi, Z.~Wang, E.D.~Walter, X.~Zhang, H.~Zhang, X.S.~Li,
A.~Tuladhar, M.~Bowden, H.-F.~Wang, S.B.~Clark, and K.M. Rosso,
{\it J.~Phys.~Chem.~C} {\bf 124}, 5275 (2020).

\bibitem{solid}
K.~Leung and A.~Leenheer, {\it J.~Phys.~Chem.~C} {\bf 119}, 10234 (2015).

\end{references}
\end{document}